\documentclass[%
 aip,
 superscriptaddress,
 amsmath,amssymb,
 reprint,%
]{revtex4-2}

\usepackage{graphicx}
\usepackage{dcolumn}
\usepackage{bm}
\usepackage{subcaption}
\usepackage[utf8]{inputenc}
\usepackage[T1]{fontenc}
\usepackage{mathptmx}
\usepackage{etoolbox}
\usepackage{xspace}
\usepackage{float}
\usepackage{titlesec}

\titlespacing*{\section}
{0pt}{1ex plus 1ex minus .1ex}{1ex plus .1ex}
\titlespacing*{\subsection}
{0pt}{1ex plus 1ex minus .1ex}{1ex plus .1ex}

\usepackage{caption}
\makeatletter
\def\@email#1#2{%
 \endgroup
 \patchcmd{\titleblock@produce}
  {\frontmatter@RRAPformat}
  {\frontmatter@RRAPformat{\produce@RRAP{*#1\href{mailto:#2}{#2}}}\frontmatter@RRAPformat}
  {}{}
}%
\makeatother

\begin{document}

\preprint{AIP/123-QED}

\title[]{Edge Magnetoplasmon Dispersion and Time-Resolved Plasmon Transport in a Quantum Anomalous Hall Insulator}
\author{Luis A. Martinez}
\affiliation{%
Lawrence Livermore National Laboratory, Livermore, CA 94550, USA.
}%
\author{Gang Qiu}
\affiliation{%
Department of Electrical and Computer Engineering, University of California, Los Angeles, Los Angeles, CA 90095, USA.
}%

\author{Peng Deng}
\affiliation{%
Department of Electrical and Computer Engineering, University of California, Los Angeles, Los Angeles, CA 90095, USA.
}%

\author{Peng Zhang}
\affiliation{%
Department of Electrical and Computer Engineering, University of California, Los Angeles, Los Angeles, CA 90095, USA.
}%

\author{Keith G. Ray}
\affiliation{%
Lawrence Livermore National Laboratory, Livermore, CA 94550, USA.
}%

\author{Lixuan Tai}
\affiliation{%
Department of Electrical and Computer Engineering, University of California, Los Angeles, Los Angeles, CA 90095, USA.
}%

\author{Ming-Tso Wei}
\affiliation{%
Lawrence Livermore National Laboratory, Livermore, CA 94550, USA.
}%

\author{Haoran He}
\affiliation{%
Department of Electrical and Computer Engineering, University of California, Los Angeles, Los Angeles, CA 90095, USA.
}%

\author{Kang L. Wang}
\affiliation{%
Department of Electrical and Computer Engineering, University of California, Los Angeles, Los Angeles, CA 90095, USA.
}%

\author{Jonathan L DuBois}
\affiliation{%
Lawrence Livermore National Laboratory, Livermore, CA 94550, USA.
}%

\author{Dong-Xia Qu}%
  \homepage{Electronic mail: qu2@llnl.gov.}
\affiliation{%
Lawrence Livermore National Laboratory, Livermore, CA 94550, USA.
}%

\date{\today}

\begin{abstract}
A quantum anomalous Hall (QAH) insulator breaks reciprocity by combining magnetic polarization and spin-orbit coupling to generate a unidirectional transmission of signals in the absence of an external magnetic field. Such behavior makes QAH materials a good platform for the innovation of circulator technologies. However, it remains elusive as to how the wavelength of the chiral edge plasmon relates to its frequency and how the plasmon wave packet is excited in the time domain in a QAH insulator. Here, we investigate the edge magnetoplasmon (EMP) resonances in Cr-(Bi,Sb)$_2$Te$_3$ by frequency and time domain measurements. From disk shaped samples with various dimensions, we obtain the dispersion relation of EMPs and extract the drift velocity of the chiral edge state. From the time-resolved transport measurements, we identify the velocity of the plasmon wave packet and observe a transition from the edge to bulk transport at an elevated temperature. We show that the frequency and time domain measurements are well modeled by loss from the microwave induced dissipative channels in the bulk area. Our results demonstrate that the EMP decay rate can be significantly reduced by applying a low microwave power and fabricating devices of larger diameter $\ge100~\mu$m. In a $R=125~\mu$m sample, a non-reciprocity of 20 dB has been realized at 1.3 GHz, shining light on using QAH insulators to develop on-chip non-reciprocal devices. 
\end{abstract}

\maketitle

\section{\label{sec:level1}Introduction}

 Quantum Hall (QH) and quantum anomalous Hall (QAH) materials are promising candidates for construction of small footprint non-reciprocal microwave devices, owing to their ability to convert an electromagnetic wave into deep-subwavelength chiral magneto-plasmonic modes.\cite{Yu2010,Chang2013,Viola2014, Liu2016, Mahoney2017PRX} Recent experiments have demonstrated robust chiral magnetoplasmon excitations in QAH materials, paving the way to developing on-chip non-reciprocal components.\cite{Mahoney2017} However, further progress is hindered by a lack of knowledge about the EMP dispersion,\cite{wang2023theory} which is critical for tuning the plasmon resonance to develop chip-level components for quantum processors, such as cryogenic circulators and isolators. Moreover, the time-domain dynamics of plasmon transport has not been studied in a QAH system. Understanding plasmon wave packet propagation over a wide temperature range will aid device development as well as material advancement.
 
 In addition to potential quantum-classical interface applications, two-dimensional (2D) magnetoplasmons in a quantum Hall system, which exhibits gapped bulk states and gapless one-way edge states, are topologically analogous to a $p$-wave topological superconductor around zero frequency.\cite{Jin2016} Recent microwave experiments have revealed the existence of topologically protected magnetoplasmons in a 2D electron gas (2DEG) in the $0.5-10$ GHz frequency range.\cite{Jin2019} As a result of the unique topological properties of 2D magnetoplasmons, magnetoplasmon zero modes have been predicted in graphene, \cite{Pan2017, Jin2017} 2DEG,\cite{Jin2016} and QAH insulators. It is therefore of great interest from both a fundamental and a practical viewpoint to measure the dispersion relation of EMPs, identify their decay times, and understand how they dissipate in the presence of microwave radiation.

 \begin{figure}
\includegraphics[width=3 in]{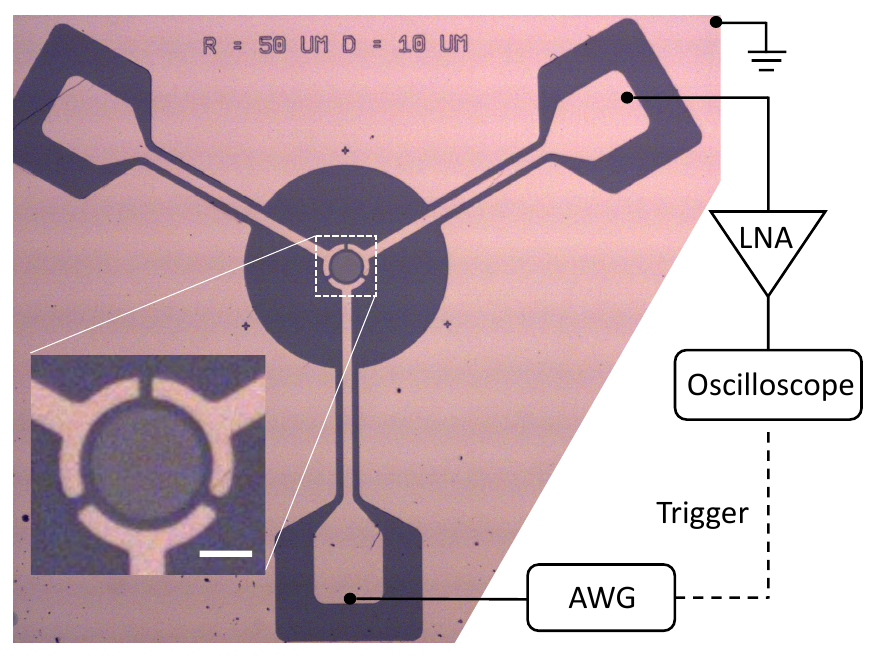}
\caption{\label{fig:Figure1}Device structure and time-resolved experimental setup. A typical device consists of a circular mesa (dark gray disk), capacitively coupled with three Au pads. The distance between the edge of the mesa and the Au pad varied from 5 to 30~$\mu$m. In the schematic of measurement circuit, the LNA represents low noise amplifier and AWG represents arbitrary wave generator.} \vspace{-10pt}
\end{figure}

\begin{figure*}
    \centering
    \includegraphics[width=1\textwidth]{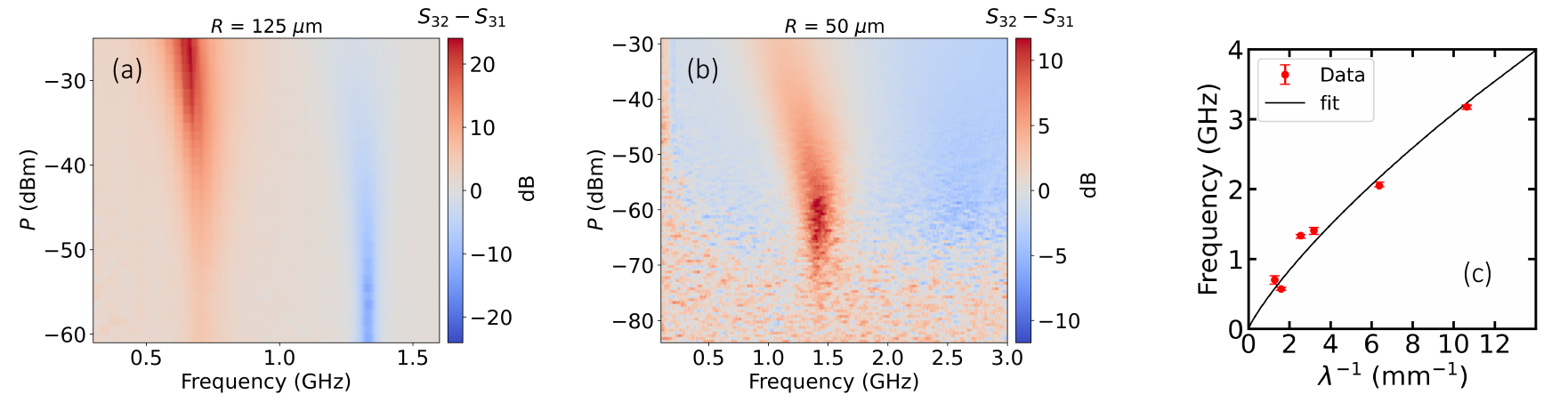}\caption{\label{fig:frequency_domain_plots}Frequency domain measurements plots of non-reciprocity $S_{32}-S_{31}$ for two selected circulator samples. The frequency response is plotted against the estimated power at the circulator port. The $R=125~\mu$m sample (a) exhibits two EMP modes, with the fundamental near 750 MHz, and the second harmonic near 1.3 GHz. The $R=50~\mu$m sample (b) exhibits a resonance near 1.5 GHz. (c) Dispersion relation of the EMP mode, obtained from the measurements on 5 samples with different radii. The solid curve is the fit to Eq. (\ref{eq:dispersion_eq}).}
    \label{fig:dispersion_plot}
\end{figure*}

Here, we report the dispersion relation and the time-resolved plasmon transport in the QAH insulator, Cr-doped (Bi,Sb)$_2$Te$_3$. Three-port circulator structures were fabricated with a mesa radius ranging from 15 to 125 $\mu$m. In the largest device, we find a 20 dB non-reciprocity for the low-frequency EMP resonance through microwave transmission measurements. In the smallest device, the non-reciprocity response still exists for the resonant frequency at 3.2 GHz, where the transverse width of EMPs is comparable to the mesa radius. Furthermore, the time-dependent measurement shows that in addition to the EMP, the bulk plasmon transport starts to appear when the temperature is increased above $\sim 1$ K. We estimated the plasmon decay time and discuss the associated dissipation mechanism.

\section{\label{sec:level1}Results}
\subsection{\label{sec:level2}Sample preparation }


The Cr$_{0.12}$(Bi$_{0.26}$Sb$_{0.62}$)$_2$Te$_3$ QAH samples with a film thickness of 6 quintuple layers (QL) were deposited using the Molecular Beam Epitaxy technique. Following the material growth, the circulator devices were patterned using a standard photo-lithography process. Peripheral co-planar waveguides made from Ti/Au 10/100 nm were deposited using an electron beam evaporator. In this work, we studied five devices with radii ranging from 15 to 125~$\mu$m (Fig. \ref{fig:Figure1}). Device dimensions are listed in the Supplemental Material (SM).
 
\subsection{\label{sec:level2}Device characterization}
Characterization of QH materials normally requires superconducting magnets with Tesla-level magnetic fields capabilities. \cite{Kumada2013} A unique feature of QAH materials is the ability to perform experiments without the need of superconducting magnets, typical off-the-shelf neodymium magnets with remnant fields on the order of tens of mT are sufficient to reach the quantized Hall state when cooling the device from room temperature. \cite{Okazaki2022} This feature allows us to perform measurements in a dilution refrigerator housing multiple qubit systems without any added noise. Details of the experimental cryogenic setup can be found in Section II of the SM.  

\subsection{\label{sec:frequency_domain_response}Microwave transmission measurements}
The frequency domain results for two selected samples are shown in Figs. \ref{fig:frequency_domain_plots} a and b. The quantity of interest is the non-reciprocity of the sample. In a perfectly reciprocal device, the non-reciprocity would be zero. The non-reciprocity is determined by the net difference in signal measured as $S_{32}-S_{31}$, when performing $S$-parameter measurements of the transmitted power on port 3 excited from port 2 ($S_{32}$) and port 1 ($S_{31}$), respectively.  For the $R=125~\mu$m sample, we observe the first EMP resonant frequency near $0.7$ GHz, and the second harmonic near $1.3$ GHz. We observed that the non-reciprocity of the first mode broadens while its frequency shifts left towards lower frequencies. Similar behaviour is observed for the second mode, with the onset at much lower port powers. The frequency response for the $R=50~\mu$m sample is shown in Fig. \ref{fig:frequency_domain_plots}b. It exhibits one EMP mode near $1.5$ GHz, with an overall magnitude of the non-reciprocity less than that of the $R=125~\mu$m sample.    

In microwave transmission measurements, the narrow EMP resonances appear as a result of the QAH phase in which the longitudinal resistance $\rho_{xx}$ is negligible and the Hall resistance $\rho_{yx}$ becomes quantized. The EMP mode in the QAH regime is the resonant edge excitation of charge density with a wavelength determined by the sample perimeter viz. $\lambda=P/n$, ($n=1, 2, ...$). $P=2\pi R$ is the perimeter of the QAH mesa, and $R$ is the radius of the mesa. The resonant frequency $f$ of the fundamental ($n=1$) and $n$th harmonic modes can be estimated to occur in the range of $0.5-10$ GHz for typical mesa parameters $P \sim 600~ \mu m$. From the EMP mode resonant frequency there follows an estimate of the phase velocity given as $v_p = fP/n$.
The dispersion relation $\omega_{EMP}(q)$ for a general 2D system with sharp edge conductance can be modeled by\cite{Volkov1988,Wassermeier1990,Petkovic2013}
\begin{equation}
f=\left[\frac{\sigma_{xy}}
{2\pi\epsilon_0\epsilon_{eff}} \left(\mathrm{ln}\frac{2}{|q|w}+1\right) + v_D\right]\lambda^{-1}
\label{eq:dispersion_eq}
\end{equation}
\noindent where $q=2\pi/\lambda$ is the wave vector, $\sigma_{xy}=C e^2/h$ the Hall conductivity, $C$ the Chern number, $\epsilon_{eff}$ the effective dielectric constant, $v_D$ the drift velocity of the 1D edge state, and $w$ the physical width of the EMP mode at the edge. 

From the microwave transmission measurements, we determined the EMP dispersion relation by plotting the dependence of the resonance frequency as a function of the inverse of the wavelength, as shown in Fig. \ref{fig:dispersion_plot}c. We fitted the experimental data using Eq. (\ref{eq:dispersion_eq}) with $\epsilon_{eff} \sim 10$, which is obtained from the finite element method simulation (Section III in the SM). The value of $\epsilon_{eff}$ is not only determined by the dielectric constant of the surrounding media, but also affected by the large-area metallic electrodes in close proximity to the mesa. The fit agrees well with the expected EMP behavior in a hard-wall edge potential, yielding $w \sim 1.4 ~\mu m$ and $v_D \sim 2.2 \times 10^4$ m/s. The extracted EMP width is smaller than $w\sim 2-5 ~\mu$m measured in a GaAs/AlGaAs 2D electron system in the presence of a strong magnetic field.\cite{Ashoori1992, Kumada2011} The value of $v_D$ is expected to be proportional with the Fermi velocity $v_D \sim 0.7v_F$ of the chiral edge state in a QH system. Therefore, the EMP dispersion measurements provide a pathway to characterize the edge state velocity that is hard to probe using spectroscopy techniques.


\begin{figure}
    \centering
    \includegraphics[width=.48\textwidth]{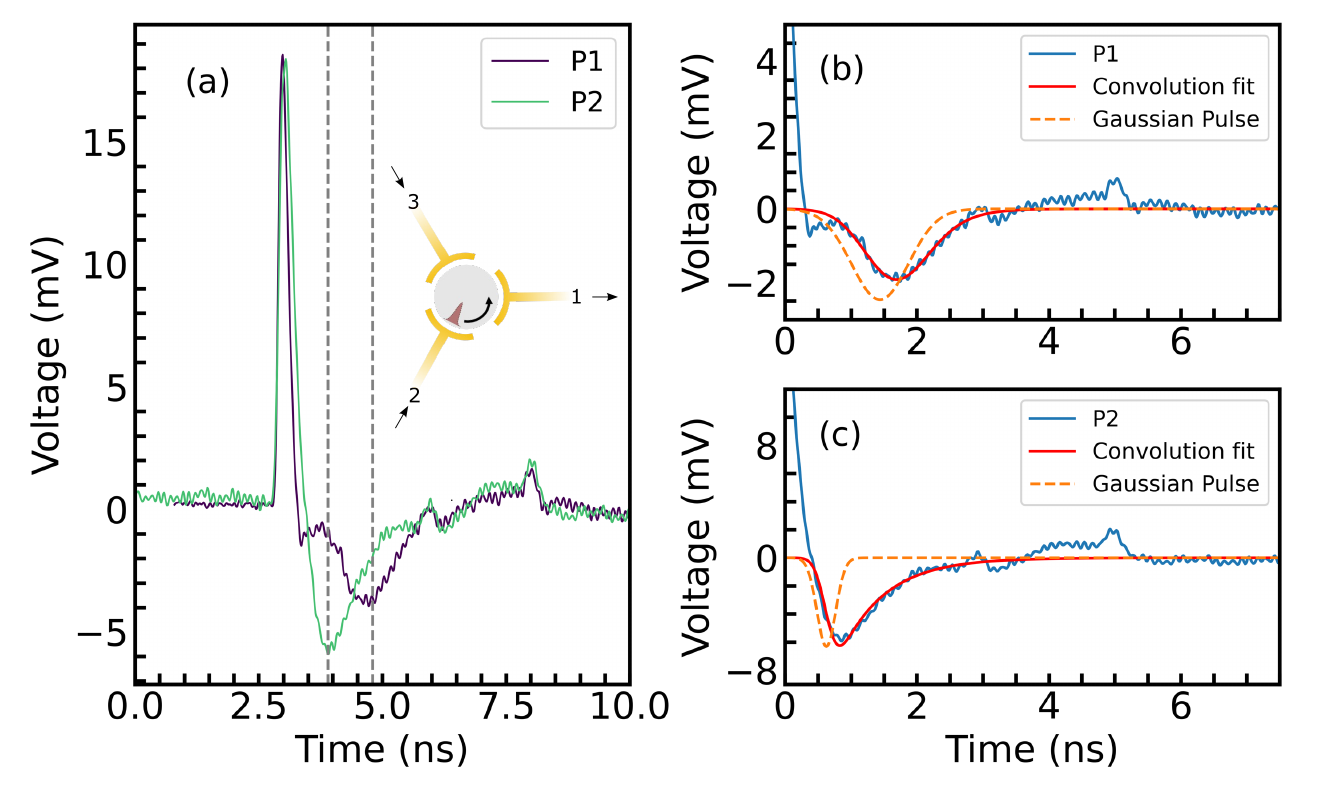}
    \caption{Time resolved plasmon response in a $R=100~\mu$m device. (a) Voltage traces as a function of time detected at port 1, when a voltage step with a magnitude of 1 V and width of 100 ns is applied to ports 2 and 3, respectively. The inset shows the port configuration and the EMP propagation in a circulator device. P1(P2) corresponds to the signal detected from port 1 and excited from port 3(2). (b, c) The enlarged EMP arrival signals and their best fits to the simulation (red). We assume the broadened EMP charge pulse that separates from the initial cross-talk peak has a Gaussian profile with a negative polarity. The Gaussian pulse obtained from the fitting is plotted as the dashed orange trace. The extreme of the Gaussian can be used to identify the plasmon propagation time. The zero time delay is determined by peak of the cross-talk signal.}
    \label{fig:TDM_plot}
\end{figure}

\subsection{\label{sec:time_domain_response}Time-resolved megnetoplasmon transport}
\begin{figure*}
    \centering
    \includegraphics[width=1\textwidth]{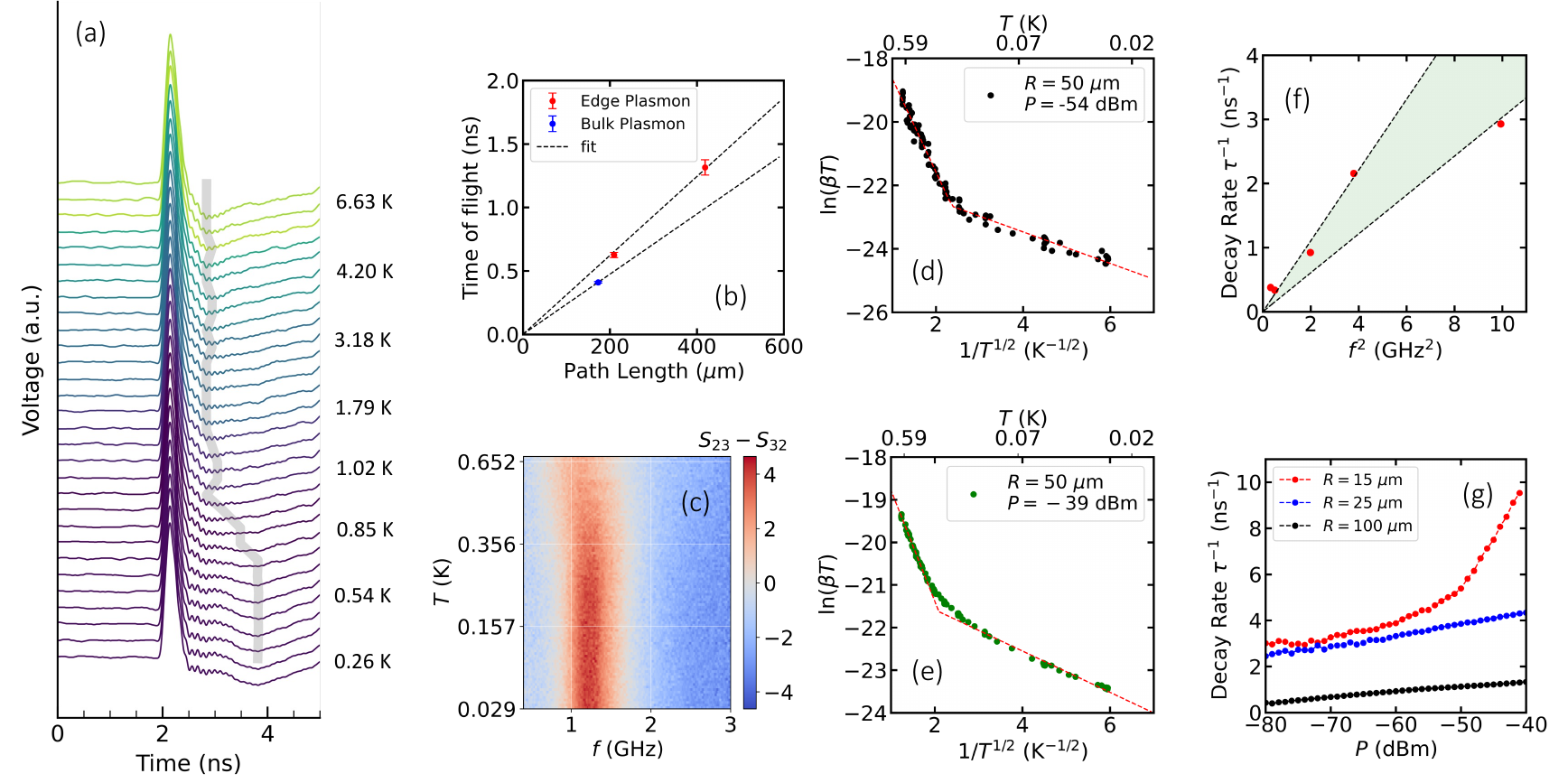}
    \caption{Temperature, microwave power, and mesa radius dependence of the edge plasmon transport. (a) Time domain measurements of the $R = 100 ~\mu$m sample with an increasing temperature for the path from port 3 to port 1. A 1 V step voltage with a rising time of 100 ps and a pulse width of 100 ns is applied to the port 3 and the signal is detected from port 1. (b) The time of flight determined at the base temperature $T=12$ mK (red dots) and the high temperature $T=6 $ K (blue dot) in the $R=100~\mu$m sample. (c) The dependence of $S_{23}$-$S_{32}$ as a function of increasing temperature with an excitation power of $-39$ dBm. (d, e) ln$(\beta T)$ versus $T^{-1/2}$ in the $R=50~\mu$m sample measured with two different input powers $P=-39$ and $-54$ dBm, respectively. The red dashed line is a piecewise linear fit to ln$(\beta T) \propto -(T_0/T)^{1/2}$. (f) The non-reciprocity decay rate $\tau^{-1}$ for the fundamental mode in all five samples as a function of $f^2$. (g) $\tau^{-1}$ as a function of microwave power for the $R=15$, $25$, and $100$ $\mu$m samples. }
    \label{fig:TDM_vs_Temp_r100_plot}
\end{figure*}

Now, let's look at the time domain results. The EMP wave packet is capacitively excited by a voltage step across the capacitor at the input port, propagating along the edge of mesa, and then capacitively coupled out through the output port capacitor. The voltage signal is amplified by a cryogenic high electron mobility transistor (HEMT) amplifier and probed by an oscilloscope as a function of time. The rise time of the voltage step varies from 100 ps to 1 ns, leading to a spatial width of the EMP wave packet in the range of $30-300~\mu m$, given that the phase velocity $v_p$ of EMP is $3\times 10^5$ m/s. Figure \ref{fig:TDM_plot}a plots the time-dependent voltage traces for the $R=100~\mu$m sample, which has a perimeter of 628 $\mu$m. The EMP wave packet was injected through the ports 2 and 3, respectively, and measured at the port 1 (inset of Fig. \ref{fig:TDM_plot}a). The measured signal consists of two distinct components, a spike with zero-time delay arising from the cross-talk through the detection circuit, and followed by a pronounced dip that shows a port-dependent delay. 


The time delay between the two dips correlates with the propagation time of the EMP over a distance of approximately $1/3$ of the perimeter along the boundary of the mesa. Using the analytical method presented in Ref. 12, we fitted the dip component with a convolution integral of a negative Gaussian function and an exponential response function $\text{exp}(-t/T_{RC})$, where $T_{RC}=0.3-0.6$ ns is the $RC$ time constant of the detection circuit (Figs. \ref{fig:TDM_plot} b and c). Here, the EMP wave packet is approximated as a negative Gaussian response due to the fact that the plasmon-transport induced dip is very close to the cross-talk spike, making the differential Gaussian profile undiscernible. Another way of extracting the time of flight of EMP propagation is described in Section IV of the SM and leads to the same result.

Next, we focus on the temperature dependence of the time domain measurements to reveal an edge-to-bulk transition. Figure \ref{fig:TDM_vs_Temp_r100_plot}a presents the voltage traces for the $R=100~\mu$m sample measured as a function of temperature $T$ from 12 mK to 6.6 K. For $T>1$ K, the edge state is no longer quantized and plasmons start to propagate in the bulk. \cite{Chang2013} As the bulk plasmon propagation undergoes a path along the mesa diameter, they appear as a shorter pulse delay compared with the path P1 along the perimeter. The time of flight for both edge and bulk plasmons is plotted in Fig. \ref{fig:TDM_vs_Temp_r100_plot}b, from which the velocity of the wave packet is determined to be $3.2 \times 10^5$ m/s and $4.2 \times 10^5$ m/s for the edge and bulk plasmons, respectively. The plasmon wave packet velocity is 10 times higher than the drift velocity $v_D$, consistent with the expected collective oscillation nature of EMPs.\cite{Kumada2013}

\subsection{\label{sec:plasmon decay}EMP decay mechanism}

Finally, we investigate the EMP decay in disk shaped QAH samples to understand the plasmon dissipation mechanism.  Figure \ref{fig:TDM_vs_Temp_r100_plot}c shows the difference response $S_{23}-S_{32}$ as a function of the frequency while sweeping the temperature from 29 mK to 0.652 K. The input port power is kept at $-29$ dBm. A decrease in frequency $f$ and the broadening of the resonant bandwidth have been observed upon increasing the temperature. We can estimate the quality factor $Q_N$ from the resonant frequency divided by the full width at half maximum ($\Delta f$) of the resonance curve: $Q_N=f/\Delta f$. The standard Lorentzian lineshape fitting to the EMP resonance allows us to obtain the non-reciprocity decay time $\tau=Q_N/\pi f$ versus $T$, as listed in Table SI. The observed effect can be explained as the result of the enhanced scattering between the EMP and the dissipative channels in the bulk region that grow with increasing temperature. 


The EMP decay mechanism in a QAH system can be quantitatively explained by the coupling of edge states to the charge puddles in the bulk. A three-port Hall circulator model \cite{Viola2014, Mahoney2017PRX} has been used to simulate the EMP decay in a QH system, with a decay rate given by $\tau^{-1} = \alpha f^2 + \beta(T)$, where $\alpha$ is a constant and $\beta (T)$ follows the variable range hopping (VRH) law $\beta \propto \frac{1}{T} \text{exp}[-(T_0/T)^{1/2}]$. \cite{Kumada2014} Because a QAH system possesses a similar topology and bulk-boundary correspondence as a QH system,\cite{Kawamura2017} we apply the same model to analyze the temperature dependence of the non-reciprocity decay rate, which is roughly proportional to the EMP decay rate (Fig. S14 c). \cite{Kumada2014} As shown in Figs. \ref{fig:TDM_vs_Temp_r100_plot} d and e, the $\text{ln}(\beta T)$ versus $1/T^{1/2}$ data shows a linear regime with a steep gradient, followed by a shallow linear regime after a threshold value. For a low power $P=-54$ dBm, the best fit gives a characteristic temperature $T_0 \sim 8.6$ K and $0.2$ K above and below the threshold temperature $T_{\text{th}}=178$ mK, respectively. For higher $P=-39$ dBm, the best fit yields $T_0 \sim 3.7$ K and $0.2$ K above and below $T_{\text{th}}=215$ mK, respectively. Raising $P$ leads to a lower $T_0$ for the VRH behavior for $T > T_{\text{th}}$, showing that the microwave radiation enhances the hopping between dissipative channels that are the main source of decay when the temperature is below 0.7 K. \cite{Lippertz2022}


To further confirm the EMP decay mechanism, we plot the $\tau^{-1}$ versus $f^2$ for various samples in Fig. \ref{fig:TDM_vs_Temp_r100_plot}f. The best fit gives $\alpha \approx 3.0 \sim 5.5\times 10^{-10}$ and $\beta \sim 0$ at low $T$. This $\alpha$ value is one order of magnitude higher than that of graphene.\cite{Kumada2014} From the simplified circuit model including the coupling capacitance $C_{loc}$ in series with the dissipation resistance $R$, the $\alpha$-related term is given by $\tau^{-1} \propto R (C_{loc} f)^2/(C_{edge} + C_{loc})$, where $C_{edge}$ is the channel capacitance \cite{Kumada2014} (also see details of the circuit model simulation in Section VII of SM). The observed $f^2$ dependence of $\tau^{-1}$ indicates the capacitively and resistively coupled dissipation mechanism exists in the QAH insulator. In addition to the large $\alpha$ value, we observed an evolution in $\tau^{-1} (P)$ with mesa radius $R$ upon microwave excitation. As shown in Fig. \ref{fig:TDM_vs_Temp_r100_plot}g, the non-reciprocity decay rate increases linearly in the $R=100$ and $25~\mu$m samples, whereas in the $R=15~\mu$m sample, $\tau^{-1}$ increases exponentially in the high power regime, suggesting that a non-linear power-dependent scattering mechanism is introduced when the mesa radius is reduced.

\section{\label{sec:level1}Discussion and Conclusions}
For a QAH system, an insulating bulk state is required for the protection of topological edge states from scattering between opposite edges. In our 6-QL Cr-doped BST sample, the QAH activation gap $120$ $\mu$eV is more than 10 times higher than the applied microwave photon energy, e.g., 4 $\mu$eV at $f=1$ GHz.\cite{Qiu2022} However, transport measurements have reported the existence of tens-of-nm size isolated charge puddles arising from the large number of superparamagnetic domains in the bulk region of the sample. \cite{Lippertz2022, Qiu2022} At high enough microwave excitation, the hopping between the neighboring puddles can possibly create additional conductive channels that are capacitively and resistively coupled to EMPs. The number of microwave excited conductive puddles is expected to be proportional to the microwave power. As a result, the characteristic temperature for the hopping between puddles will decrease with increasing $P$. This is consistent with our observation that $T_0$ is suppressed with a higher microwave excitation for $T > T_{\text th}$. As the device radius decreases, the smaller device has a greater chance of forming percolating paths that allow edge-to-edge scattering, due to the shorter edge-to-edge distance and the higher resonant frequency, or equivalently, the higher photon energy. This effect is reflected in the decay rate $\tau^{-1}(P)$ as being dramatically enhanced in the smallest sample. 

In summary, we have studied the dispersion relation of EMPs and their decay in a QAH insulator with microwave transmission measurements and time-resolved measurements. We find the transverse width $\sim 1~\mu$m of EMP is much wider than the edge channel width reported in the equilibrium transport measurement, which is below tens of nanometers.\cite{Qiu2022} The wide EMP wave packet width indicates that the EMPs differ from their equilibrium counterparts by the presence of electromagnetic fields that influence the local potentials at the edge. We confirmed the collective nature of the EMPs by showing that the EMP wave packet's propagation velocity is one order of magnitude higher than its drift velocity. The microwave power dependent decay rate is featured with a characteristic hopping temperature, indicating that VRH is still the dominant dissipation mechanism upon the microwave radiation. From the size and power dependence of the microwave transmission measurements, the non-reciprocity decay time is found to scale with the mesa radius, which points to the enhanced scattering between macroscopically separated edge states as the device shrinks to tens of micrometers. Our results will serve to inspire investigation into designing new device structures, toward the realization of low-loss chiral interconnects based on QAH insulators.

\begin{acknowledgments}
We would like to thank Cianpaolo P. Carosi for helpful discussions, and Kristin Beck for assistance in performing the experiments. This work was performed under the auspices of the US Department of Energy by Lawrence Livermore National Laboratory under Contract No. DE-AC52-07NA27344. The project was supported by the Laboratory Directed Research and Development (LDRD) programs of LLNL (21-ERD-014).The authors also acknowledge support from NSF Convergence Accelerator program under Grant 2040737.
\end{acknowledgments}

\nocite{*}
\bibliography{MainRef}

\end{document}


\preprint{AIP/123-QED}

\title[]{Supplementary Information: Edge Magnetoplasmon Dispersion and Time-Resolved Plasmon Transport in a Quantum Anomalous Hall Insulator}
\author{Luis A. Martinez}
\affiliation{%
Lawrence Livermore National Laboratory, Livermore, CA 94550, USA.
}%
\author{Gang Qiu}
\affiliation{%
Department of Electrical and Computer Engineering, University of California, Los Angeles, Los Angeles, CA 90095, USA.
}%

\author{Peng Deng}
\affiliation{%
Department of Electrical and Computer Engineering, University of California, Los Angeles, Los Angeles, CA 90095, USA.
}%

\author{Peng Zhang}
\affiliation{%
Department of Electrical and Computer Engineering, University of California, Los Angeles, Los Angeles, CA 90095, USA.
}%

\author{Keith G. Ray}
\affiliation{%
Lawrence Livermore National Laboratory, Livermore, CA 94550, USA.
}%

\author{Lixuan Tai}
\affiliation{%
Department of Electrical and Computer Engineering, University of California, Los Angeles, Los Angeles, CA 90095, USA.
}%

\author{Ming-Tso Wei}
\affiliation{%
Lawrence Livermore National Laboratory, Livermore, CA 94550, USA.
}%

\author{Haoran He}
\affiliation{%
Department of Electrical and Computer Engineering, University of California, Los Angeles, Los Angeles, CA 90095, USA.
}%

\author{Kang L. Wang}
\affiliation{%
Department of Electrical and Computer Engineering, University of California, Los Angeles, Los Angeles, CA 90095, USA.
}%

\author{Jonathan L DuBois}
\affiliation{%
Lawrence Livermore National Laboratory, Livermore, CA 94550, USA.
}%

\author{Dong-Xia Qu}%
\affiliation{%
Lawrence Livermore National Laboratory, Livermore, CA 94550, USA.
}%

\date{\today}


\maketitle

\section{\label{sec:level1}Sample preparation}
\begin{figure*}[h]
\includegraphics[width=3 in]{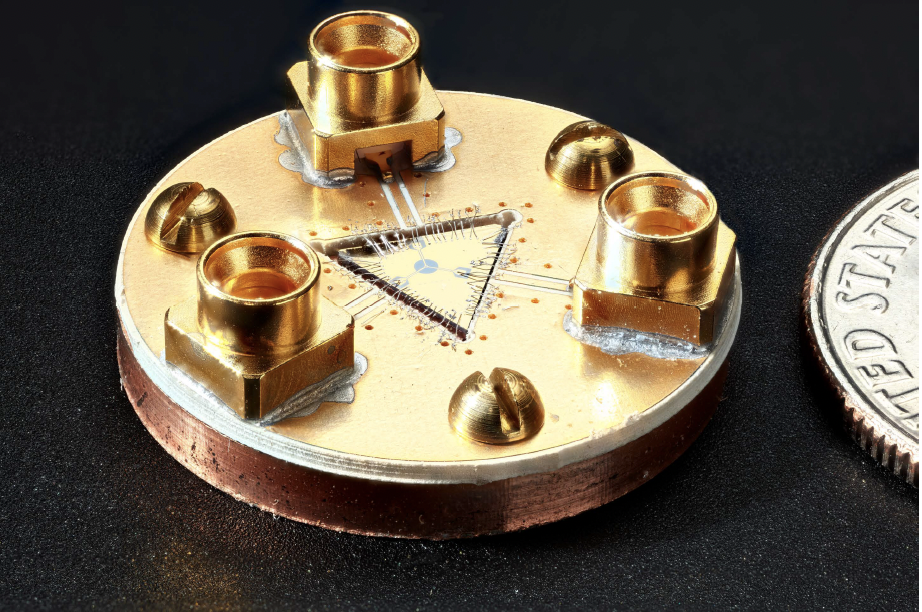}
\caption{\label{fig:device_image} Photograph of the circulator device mounted on a printed circuit board. U.S. dime coin is shown for scale.}
\end{figure*}

\begin{table*}[h]
\caption{Parameter list for circulator devices. 1st and 2nd $f$ correspond to the fundamental and the second-harmonic magnetoplasmon (EMP) frequency, respectively. $v_p$ is the EMP phase velocity. The non-reciprocity quality factor $Q_N=f /\Delta f$ is obtained from the Lorentzian fit, where $\Delta f$ is the full width at half maximum (FWHM). The corresponding non-reciprocity decay time is given as $\tau = Q_N/\pi f$. The quality factor $Q_{\text{3dB}}$ and EMP decay time $\tau_{\text{3dB}}=Q_{\text{3dB}}/\pi f$ can also be estimated from the width at $-3$ dB of the resonance peak. The table lists the highest quality factor obtained using two methods. $\epsilon_{eff}$ refers to the effective dielectric constant, described in Section \ref{sec:dielectric_sim}.}
\centering
\vspace{5pt}
\setlength{\tabcolsep}{3pt}
\begin{ruledtabular}
\begin{tabular}{l*{10}{c} r}
Sample       & Radius   & Gap & 1st $f$ & 2nd $f$ & $\nu_{p}$ & $Q_N$ & $\tau$  & $Q_{\text{3dB}}$& $\tau_{\text{3dB}}$& $\epsilon_{eff}$ \\
      & ($\mu$m)  &  ($\mu$m)& (GHz)&  (GHz) &  ($10^5$ m/s) & &  (ns) &  &  (ns) & for 1st $f$ / 2nd $f$\\
\hline
C4      &125  & 30 & 0.7 & 1.3 & 5.5 & 7 & 3.2 & 19 & 8.6 & 8.64 / 8.35\\
C9       &100  & 15 & 0.6 & - & 3.7 &  5 & 2.5 & 13 & 6.9 & 9.45 / 9.20\\
C10      &50   & 10 & 1.4 & - & 4.5 & 4 & 0.94 & 7 &  1.6 & 10.08 / 9.44\\
C14  &25   & 10 & 2.1 & - & 3.1 & 2 & 0.32 &-&- & 9.88 / 8.89\\
C15  &15   & 5 & 3.2 & - & 3.0 & 3 & 0.29  &-&- & 11.26 / 9.87

\vspace{0pt}
\end{tabular}
\end{ruledtabular}
\label{Table:Sample_Parameters}
\end{table*}

The quantum anomalous Hall (QAH) samples were deposited using Molecular Beam Epitaxy. The base pressure of the Perkin-Elmer chamber was kept below $10^{-9}$ Torr. Semi-insulating GaAs (111) substrates were pre-treated at 600 $^\circ C$ under a Te-rich environment. High-purity Cr(99.995\%), Bi(99.999\%), and Te(99.9999\%) sources were evaporated from Knudsen effusion cells while Sb(99.999\%) was evaporated from a cracker cell. The Cr, Bi, Sb, and Te cell temperatures were kept at 1080 $^\circ C$, 472 $^\circ C$, 372 $^\circ C$, and 340 $^\circ C$, respectively, whereas the substrate was heated to 200 $^\circ C$. Bright streaky features were observed from in-situ real-time high-energy electron diffraction (RHEED) patterns, confirming the formation of highly crystalline QAH 2D films.

Following the material growth, the circulator devices were patterned using a standard photo-lithography process. Peripheral co-planar waveguides made from Ti/Au 10/100 nm were deposited using an electron-beam evaporator. In this work we studied five devices with radii ranging from 15 to 125~$\mu$m. A device image is shown in Fig. S\ref{fig:device_image} and the device dimensions are listed in the Table S\ref{Table:Sample_Parameters}.

\section{\label{sec:level1}Experimental Configuration}
\begin{figure*}[h]
\includegraphics[width=3.5 in]{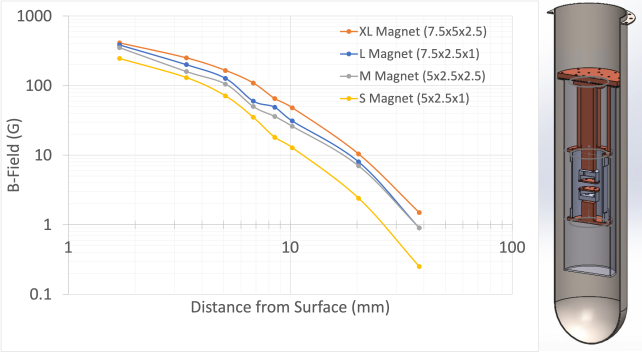}
\caption{\label{fig:magnet_characterization} Magnetic field strength at room temperature of various neodymium magnets. Field measured along the axial direction. Right panel illustrates the cross-section of the double magnetically shielded sample holder used to carry out the experiments in a dilution refrigerator.}
\end{figure*}

The experimental configuration was designed with dual cryogenic magnetic shields to contain the magnetic field within the desired sample space. This avoids contaminating the entire sample space on the mixing chamber plate with unwanted magnetic fields. The permanent neodynium magnet rings were characterize along the axial direction at room temperature with a Lakeshore 455 Gaussmeter and transverse Hall probe, see Fig. S\ref{fig:magnet_characterization}. Note, this configuration makes it possible to measure the QAH samples without the use of large superconducting magnets, and can be done concurrently in environments housing superconducting qubit experiments. 

\subsection{Frequency Domain measurements}{\label{subsubsec:symmetric_FDM}}

\begin{figure*}[h]
\includegraphics[width=3.5 in]{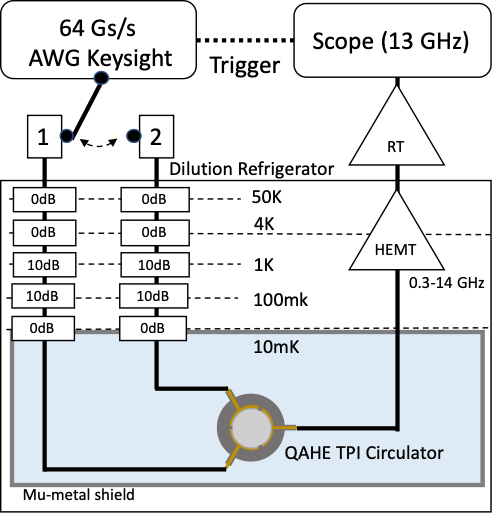}
\caption{\label{fig:Exp_Configuration} A 64 Gs/S AWG generates pico-second rise time pulses for the time-resolved transport measurement. Time domain traces are captured with a 13 GHz oscilloscope triggered by the AWG.}
\end{figure*}

A direct measurement of non-reciprocity was achieved by obtaining $S_{21}$ transmission curves for two input ports and one output port of the circulator. The output port (e.g., port 3) is connected to a wide-band HEMT followed by a room temperature wide-band amplifier. The wiring configuration is analogous to Fig. S\ref{fig:Exp_Configuration} with the scope and AWG replaced with a Vector Network Analyzer (VNA). In this case subtracting the two $S_{21}$ measurements gives the non-reciprocity between the two input ports, port 1 and port 2, and the output port, port 3 of the circulator:
\begin{equation}
\Gamma_{21}-\Gamma_{12} = S_{31} - S_{32}
\label{eq:non_reciprocity_typ}
\end{equation}


The symmetric measurement procedure for extracting the reflection and transmission parameters of a non-reciprocal device inside a dilution refrigerator operating at temperatures below 10 mK is summarized next. The RF properties learned from the $S$-parameters of a device under test (DUT) are typically the quantities of interest. At room temperature the $S$-parameters are easily measured directly with a VNA. However, when the DUT is placed in a dilution refrigerator the extra components complicate the measurements of the $S$-parameters. This is because measurements in dilution refrigerators require several feet of SMA cabling, starting at room temperature and reaching to the mixing chamber plate. The input lines typically are loaded with a series of attenuators, whereas the output lines are loaded with amplifiers and other passive non-reciprocal RF components (e.g., isolators). 

\begin{figure}[h]
   \centering
   \includegraphics[width = .4\textwidth]{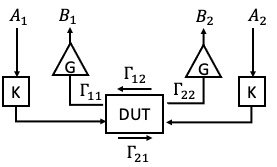} 
   \caption{Measurement schematic for extracting the reflection and transmission coefficients for the device under test (DUT). The in-going circuit, including attenuation, is lumped together and its properties are summarized by the lumped element $K$. Similarly, the out-going circuit is summarized with the parameter $G$, which includes attenuation and gain from the amplifiers. It is assumed that the circuit is symmetric.}
   \label{fig:diagram}
\end{figure}

Extracting the $S$-parameters of the DUT is summarized in Fig. S\ref{fig:diagram} which illustrates a symmetric measurement configuration for a 2-port device. Note, in practice a directional coupler is used on each port to direct the signals accordingly. The components denoted by $K$ include the input attenuation right up to the DUT (i.e., for signals traveling towards the DUT ports). Similarly, the components denoted by $G$ include the output transmission properties for the signals traveling away from the ports which also includes any amplification. 

Consider an input signal $V^+$ at input port $A_1$. After the attenuation the signal amplitude is $V^+/K$, and some portion of the signal is reflected at the input port of the DUT. Let's denote the reflection by $\Gamma_{11}$, then the reflected signal is $V^+\Gamma_{11}/K$. Next, the reflected outgoing signal is amplified and we have 
\begin{equation}
V^- = \frac{V^+\Gamma_{11}G}{K},
\end{equation}
where $G$ is the gain factor (which encompasses any attenuation along the output line). As measured by the VNA, in an $S_{21}$ measurement between ports $A_i$ and $B_i$ for $i=1$, we obtain the following relationship
\begin{equation}
S_{21} = \frac{V^-}{V^+} = \frac{\Gamma_{11}G}{K}.
\end{equation}

In a similar fashion we can arrive at the other port $S$-parameters for the DUT. Converting to logarithmic units via $20\log(V^-/V^+)$ we find a set of four equations for the port parameters of the DUT.
\begin{align}
&\Gamma_{11} = SB_1A_1 - G +K ~ \text{[dB]} \\
&\Gamma_{22} = SB_2A_2 - G + K ~ \text{[dB]}\\
&\Gamma_{21}= SB_2A_1 - G + K~ \text{[dB]} \\
&\Gamma_{12} = SB_1A_2 - G + K~ \text{[dB]},
\end{align}
where we use the notation $SB_iA_i $ to denote the VNA measurements as illustrated in Fig. S\ref{fig:diagram}. Note that the line G and K can be measured independently and thereby arrive at an estimate of the actual DUT's $S$-parameters. An approximation can be made if $G$ and $K$ are measured at room temperature instead of cryogenic temperatures. Notice that this is equivalent to removing network elements not part of the DUT, which effectively moves the measurement plane to the ports of the DUT. In the symmetric case (gain $G$ and attenuation $K$ are identical) we can subtract out the gain and attenuation factors, i.e., 
\begin{equation}
\Gamma_{21}-\Gamma_{12} = SB_2A_1-SB_1A_2,
\label{eq:non_reciprocity}
\end{equation}
which gives the non-reciprocity of the DUT.

\subsection{Time Domain Measurements}
The experimental configuration for the time domain measurement is summarized in Fig. S\ref{fig:Exp_Configuration}. The experiments were carried out at the mixing chamber plate of a dilution refrigerator having a base temperature below 10 mK. The sample space was enclosed with dual magnetic shields, and a total of 20 dB of attenuation was used on each input line. The output line was equipped with a wideband LNF $0.3-14$ GHz HEMT pre-amplifier housed at the 4K plate which was followed by an additional wideband room temperature amplifier. The time domain signals were generated with a 64 Gs/S M8195a arbitrary waveform generator (AWG), and the time traces were recorded with a MSOX 9130a 13 GHz oscilloscope. The corresponding monitor channel (180 deg out of phase from main signal) of the AWG was used as a trigger for the oscilloscope.

\section{\label{sec:dielectric_sim}Effective Dielectric Constant Simulation}

\begin{figure*}
\includegraphics[width=4 in]{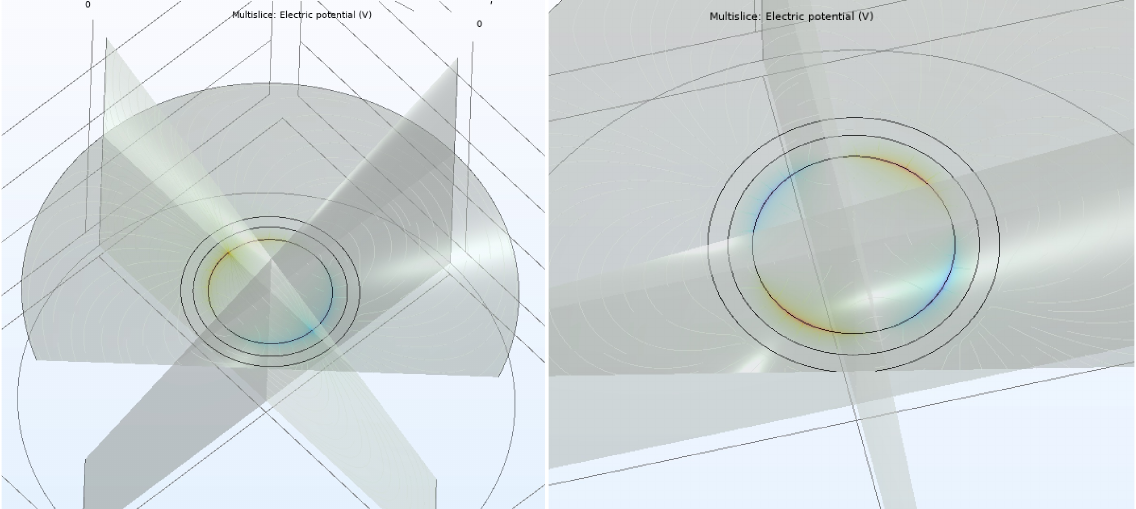}
\caption{Simulations of the electric potential created by the 1st (left) and 2nd (right) modes of a magnetoplasmon excitation in the topological circulator device. These simulations are used to calculate the effective dielectric constant.}
\label{fig:dielectric_calc}
\end{figure*}

The geometry of the magnetoplasmon device and the proximity to metallic electrodes can influence the effective dielectric constant.\cite{grodnensky1991nonlocal,ashoori1992edge} To calculate the value of the effective dielectric constant to use in the equation for the magnetoplasmon dispersion, we employ finite element analysis models in COMSOL that match the device geometries and materials. We employ the quasi-static approximation, as the dimensions of the device relevant to the effective dielectric constant are much smaller than the distance light travels on the timescale of the GHz excitation. For these models the substrate is GaAs with a dielectric constant of 12.5. On this substrate is the Cr-doped (Bi,Sb)$_2$Te$_3$ material, 6 nm thick, with a dielectric constant of 65.\cite{yin2017plasmonics} The Ti/Au contacts surrounding the Cr-doped (Bi,Sb)$_2$Te$_3$ disc are represented by an annulus that is 100 nm thick with a dielectric constant of $2.5 \times 10^5$.\cite{pandey2016non} The vacuum region above the circulator has a relative permittivity of 1. The diameter of the Cr-doped (Bi,Sb)$_2$Te$_3$ disc and the inner and outer diameters of the Ti/Au annulus depend on the particular device geometry. The magnetoplasmon excitation is simulated with a charge on the circumference of the Cr-doped (Bi,Sb)$_2$Te$_3$ disc that varies as a sinusoid with 1 or 2 periods, wrapped around the disc, to represent either the 1st or 2nd mode of the circulator, see Fig. S\ref{fig:dielectric_calc}. Once the electric fields in the volumes of the device and vacuum above are calculated we integrate the dielectric constant over all simulation space, weighted by the electric field magnitude. After normalizing, we arrive at an effective dielectric constant for a particular circulator device geometry, shown in Table S\ref{Table:Sample_Parameters}.

\begin{figure*}
\includegraphics[width=5in]{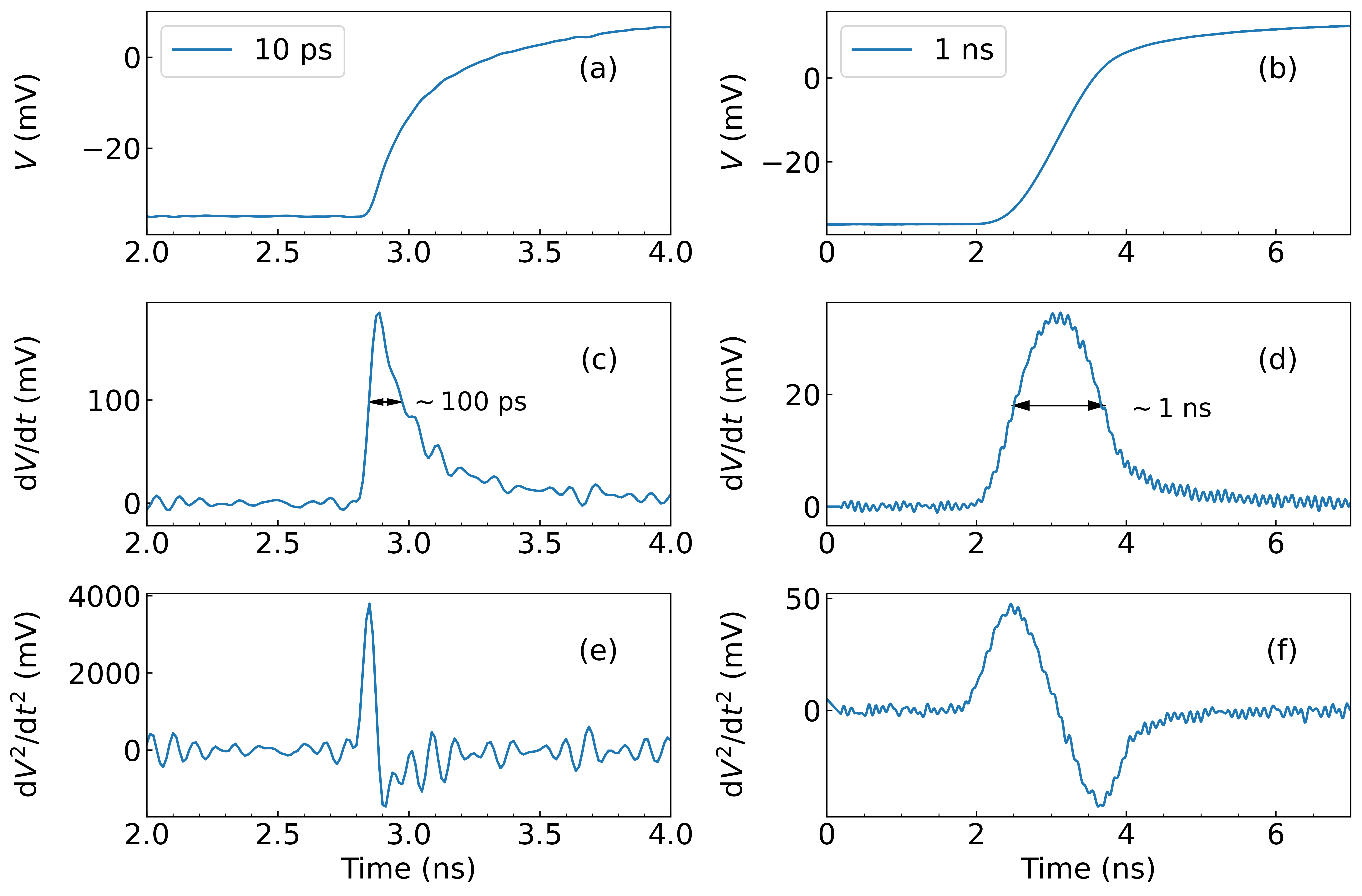}
\caption{\label{fig:Rising_Time_input}Plasmon excitation with a pulse injection. (a, b) The rise edge of the excitation voltage step measured at the input port. The rise time is set to be 10 ps (a) and 1 ns (b) at the AWG, respectively. The 10 ps excitation step is broadened to $\sim$100 ps due to the built-in RF cables and connectors in the dilution refrigerator. (c, d) The derivative of the excitation voltage step d$V$/d$t$, which corresponds to the injected pulse wave packet for the 100 ps (d) and 1 ns (d) rising edge step, respectively.  (e, f) The second derivative of the excitation voltage step d$V^2$/d$t^2$, which corresponds to the expected pulse profile detected by the oscilloscope.}
\end{figure*}

\section{\label{sec:level1}EMP Delay Time Estimated from the Time-Domain Measurements}

\subsection{\label{sec:level2}Rise Time Dependent Response}

\begin{figure*}
\includegraphics[width=5.5in]{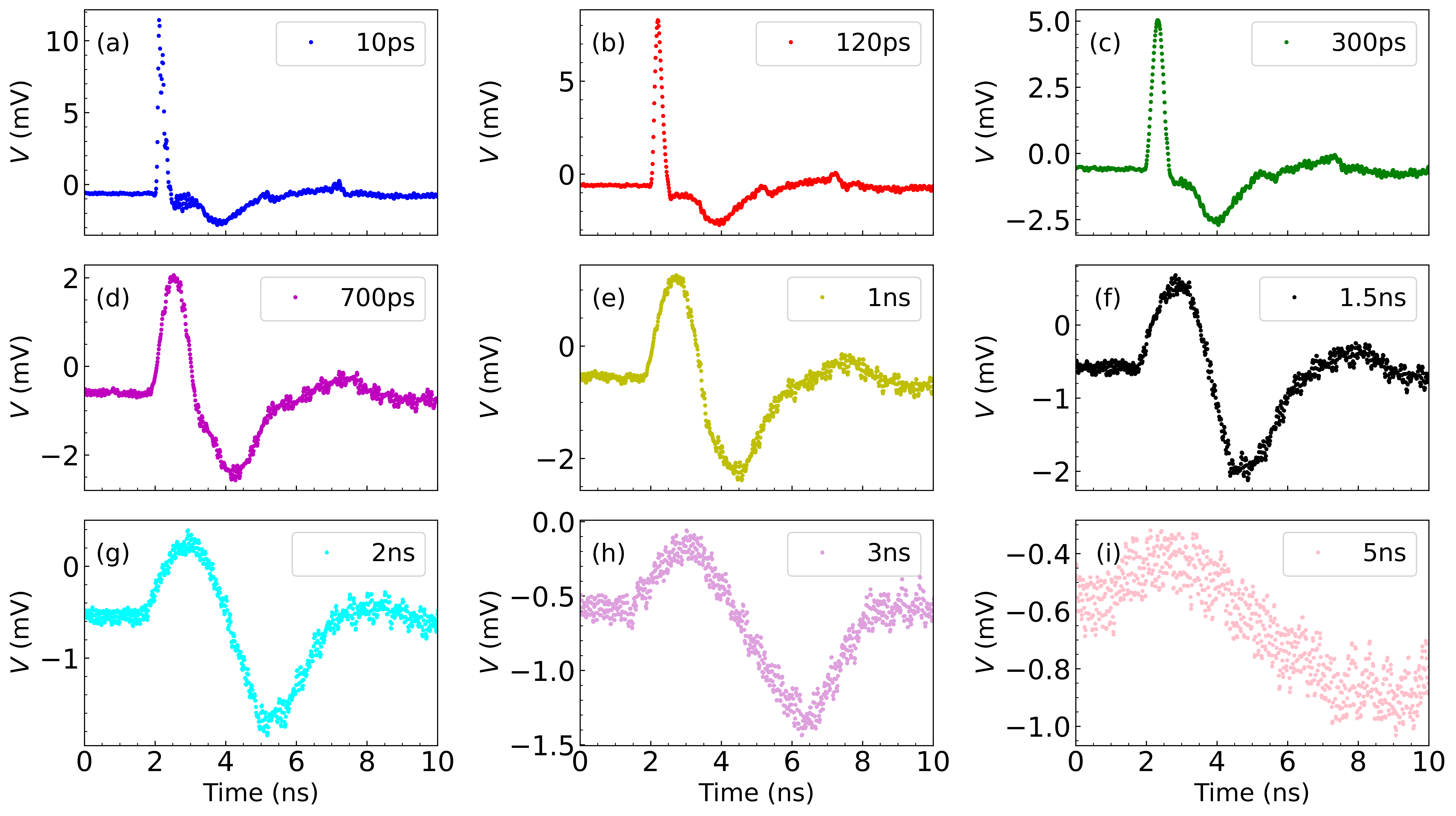}
\caption{\label{fig:Rise_Time_all}Time-resolved response for various rise time excitation. The voltage step is injected at port 3 and detected at port 1 with a path length of P1 for the $R=100~\mu$m sample.}
\end{figure*}

\begin{figure*}
\includegraphics[width=5in]{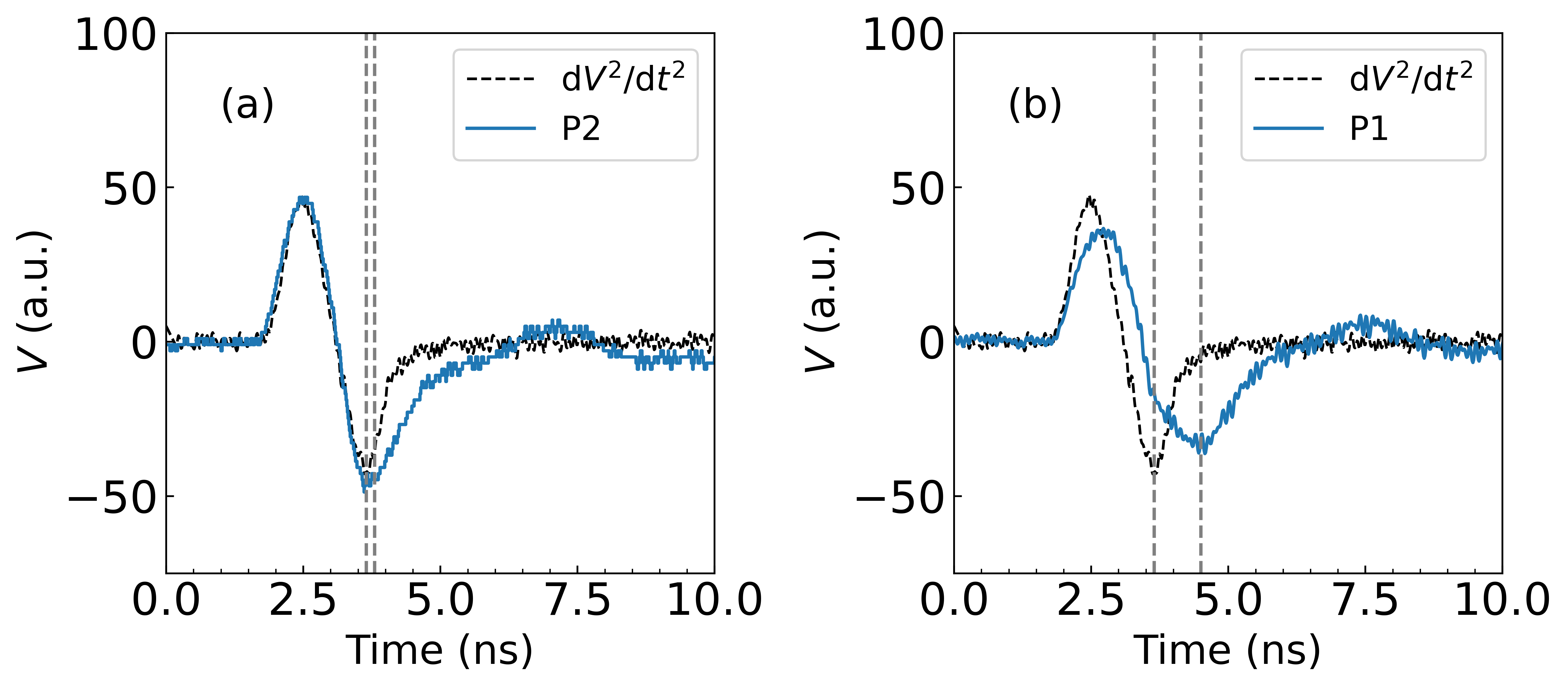}
\caption{\label{fig:Delay_Time_1ns}Time-resolved response for the 1 ns rise time excitation. (a, b) The voltage signal detected for the short path P2 (blue trace in a) and the long path P1 (blue trace in b) in comparison with the second derivative of the input voltage step (black dashed trace). The dashed lines mark the minima of each trace.}
\end{figure*}

In the time-domain measurements, the pulsed wave packet is injected and detected through the in-plane capacitive ports by applying a voltage step with a rise edge varying from 100 ps to 2 ns. The gap distance between the input transmission line and the QAH mesa is shown in Table SI for each sample. Figure S\ref{fig:Rising_Time_input} shows the direct measurements of the rising edge of the voltage step after going through all the cables and RF components in the dilution refrigerator. The 10 ps rise time voltage step generated from the AWG is broadened to approximately 100 ps by the built-in RF components of the system. As the voltage step is capacitively applied to the sample, the waveform of the induced charge pulse is proportional to the derivative of the voltage step (Figs. S\ref{fig:Rising_Time_input} c and d). The excited plasmon wave packet then propagates in the mesa and is coupled out through a symmetric capacitor at the output port. The output signal is expected to have a second order differential profile of the voltage step, as shown in Figs. S\ref{fig:Rising_Time_input} e and f.

A rise time dependence of the time domain response is observed (Fig. \ref{fig:Rise_Time_all}). In the measurement, the excitation voltage step had an amplitude of one volt and a pulse width of 100 ns. For a rise time less than 700 ps, the detected signal exhibits two components: a sharp spike with a zero-time delay followed by a dip. Here, the EMP channel acts like a slowed transmission line. Once the rise time is above 700 ps, the sharp spike vanishes. In this way, the true plasmon response can be identified when the rise time is too long to excite the high frequency spike.  As shown in Fig. S\ref{fig:Delay_Time_1ns}, the time-domain trace for the path P2 closely follows the d$V^2$/d$t^2$ plot, while the output signal for the path P1 displays a notable delay. Using this long rise time measurement, we can determine that paths P1 and P2 have a delay time of approximately 1 ns, which is consistent with the fast rise time result shown in the main text.


\subsection{\label{sec:level2}Pulse Width Dependent Response}
\begin{figure*}
\includegraphics[width=5.5in]{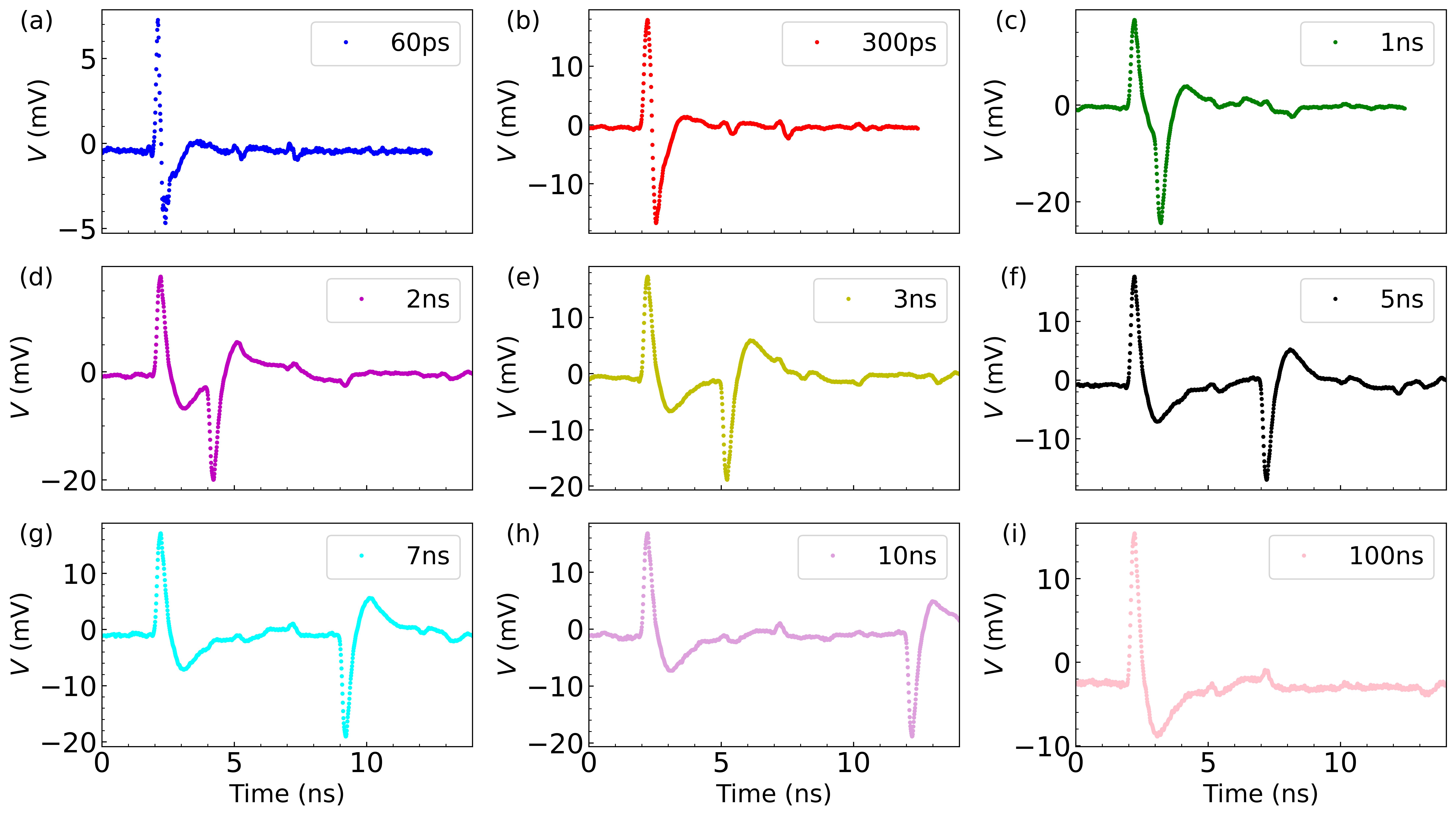}
\caption{\label{fig:Pulse_width_all}Time-resolved response for various pulse width excitation. The voltage step is injected at port 3 and measured at port 2 with a fixed rise time of 100 ps for the $R=100~\mu$m sample. }
\end{figure*}

We also measured the pulse width dependence of the time-resolved response with a fixed 100 ps rise time (Fig. S\ref{fig:Pulse_width_all}). The rising edge response cannot be distinguished from that of the falling edge when the pulse width is less than 1 ns. Rising and falling edges of the pulse exhibit an asymmetric response pattern. If the rising edge spikes followed by a dip, then the falling edge spikes followed by a peak.

\begin{figure*}[h]
\includegraphics[width=4.5 in]{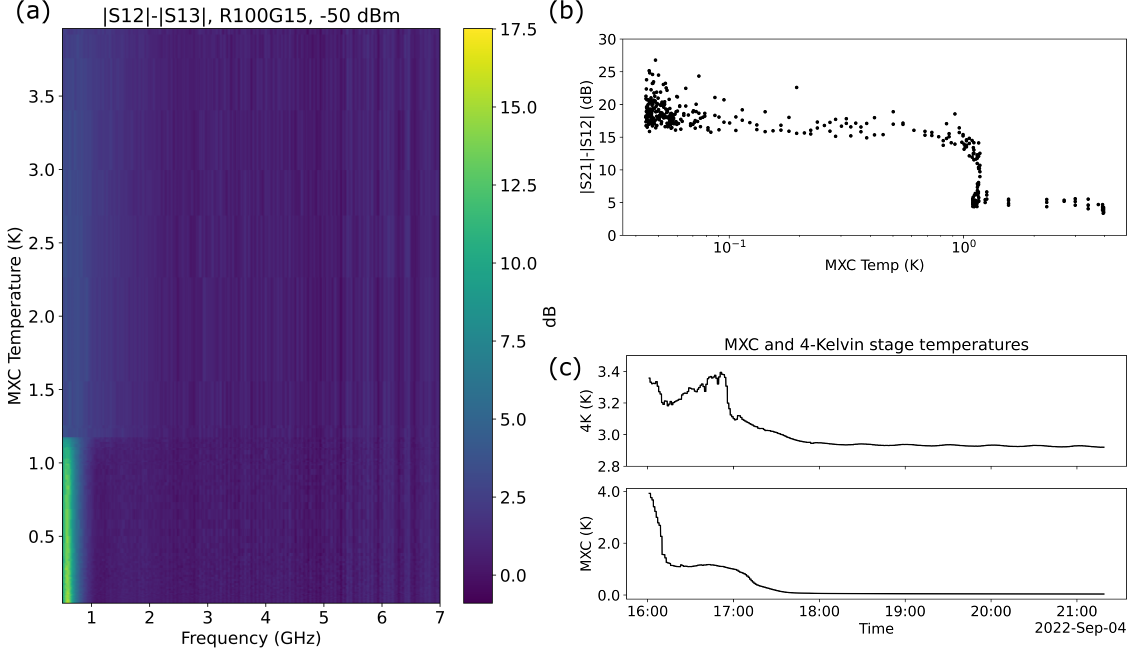}
\caption{\label{fig:FDM_vs_temp_R100} (a) Wideband frequency domain response as a function of the mix chamber temperature recorded during the cooldown cycle of the dilution refrigerator. At temperatures below 1.2 K a clear isolation signal is observed at approximately 600 MHz. (b) A transition from the bulk to edge states manifest as a reciprocal to non-reciprocal transition, and which is observed to occur between 1.2 K and 700 mK. The magnitude of the non-reciprocity plateaus at approximately 20 dB. (c) The relevant temperatures are plotted against time. Note, the HEMT amplifiers are located at the 4K stage and the observed variation in temperature is not expected to generate significant changes in gain.}
\end{figure*}

\section{\label{sec:level1}Temperature Dependent Microwave Transmission Measurements}

\subsection{\label{sec:level2}Frequency and time domain responses in the $R=100~\mu$m sample}

\begin{figure*}
\includegraphics[width=5.5in]{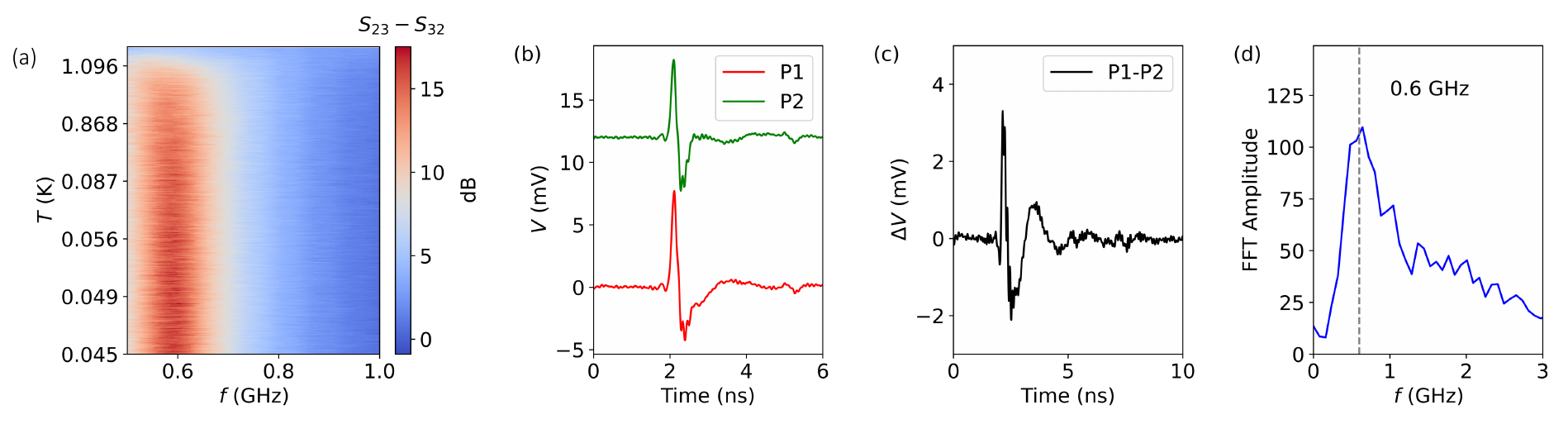}%
\caption{\label{fig:R100} Frequency and time-resolved response in the $R=100~\mu$m Sample. (a) The $S_{23} - S_{32}$ response as a function of the MXC temperature with the microwave excitation power at -70 dBm. (b) The time-domain response for two paths P1 (red) and P2 (green). (c) The difference between the P1 and P2 paths. By subtracting the two paths, we are left with the signal associated with the EMP and suppress the cross-talk signal. (d) Fast Fourier transform (FFT) of the P1-P2 signal in c. The FFT peak occurs at $\sim 0.6$ GHz, consistent with the resonance frequency in microwave transmission measurements as shown in a.}
\end{figure*}

\begin{figure*}
\includegraphics[width=5.5in]{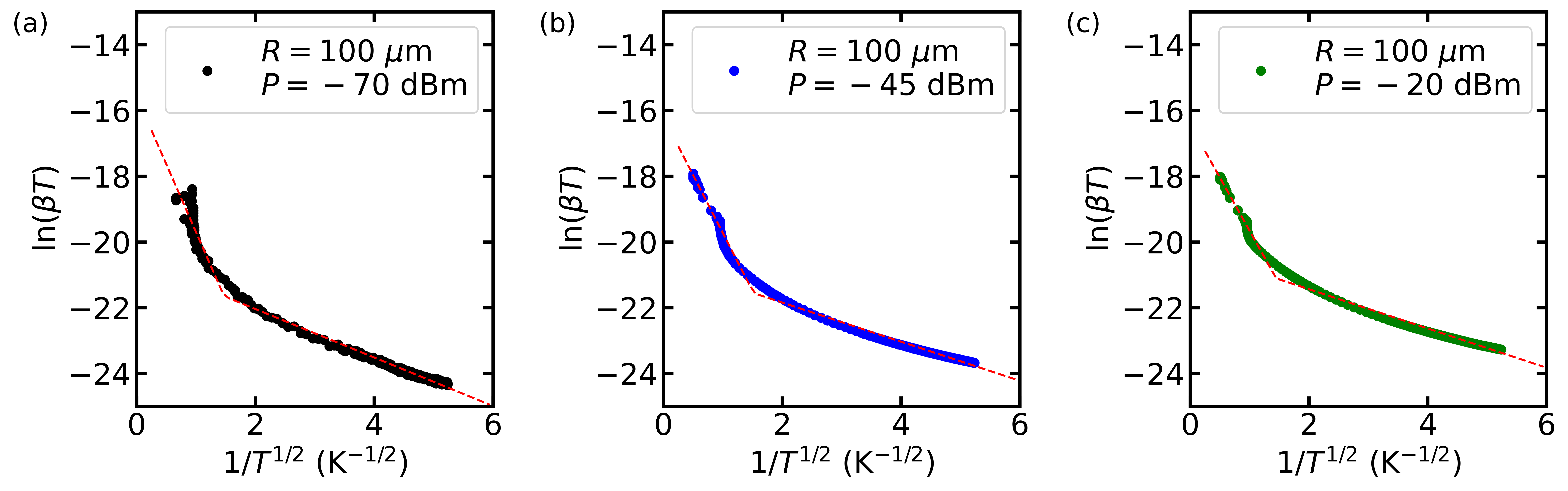}
\caption{\label{fig:R100_tau} Non-reciprocity decay time for different microwave radiation power in the $R=100~\mu$m sample. The $\text{ln}(\beta T)$ vs. $T^{-1/2}$ with the microwave excitation powers at $-70$ dBm (a), $-45$ dBm (b), and $-20$ dBm, respectively. The red dashed lines represent the piecewise linear fit.}
\end{figure*}

\begin{figure*}
\includegraphics[width=5.5in]{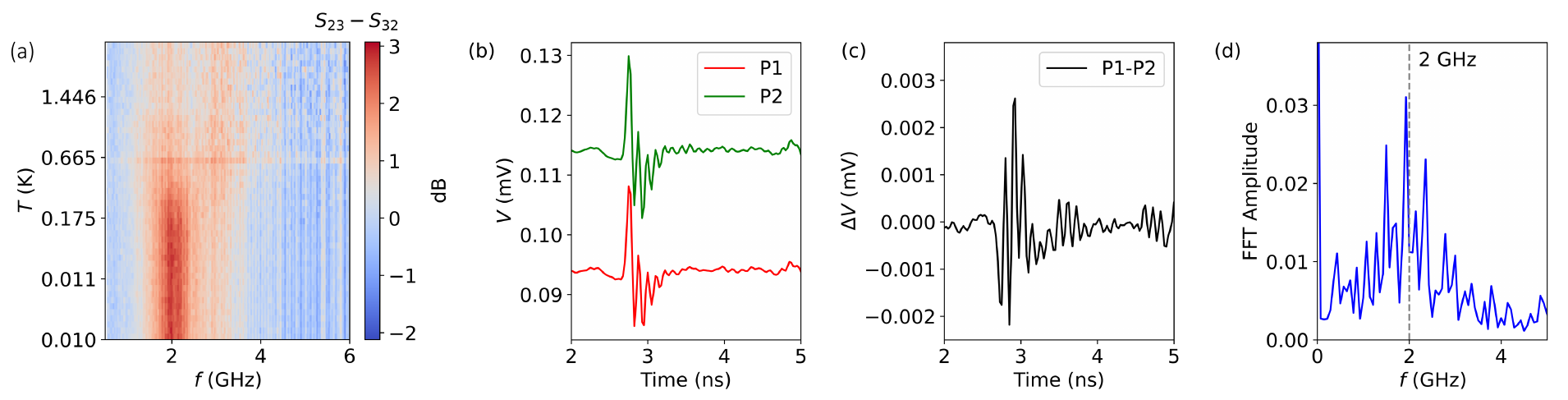}
\caption{\label{fig:R25} Frequency and time resolved response in the $R=25~\mu$m sample. (a) The $\Delta S_{23} - \Delta S_{32}$ response as a function of the MXC temperature with the microwave excitation power at $-70$ dBm. (b) The time domain response for two paths P1 (red) and P2 (green). (c) The difference between P1 and P2. (d) Fast Fourier transform (FFT) of P1-P2 in c. The FFT peak occurs at $\sim 2$ GHz, consistent with the resonance frequency in the microwave transmission measurements.}
\end{figure*}

We obtained the temperature dependant frequency domain responses of the $R=100~\mu$m sample during a cooldown cycle using the symmetric measurement configuration described in Section \ref{subsubsec:symmetric_FDM}. Using a room temperature RF switch, non-reciprocity measurements were obtained by alternating VNA transmission measurements between two input ports of the $R=100~\mu$m sample. As shown in Fig. S\ref{fig:FDM_vs_temp_R100}, a clear transition from normal reciprocal behaviour, associated with the bulk conduction states, to non-reciprocal behaviour associated with the edge states is observed at approximately 1.2 K. Further analysis tracking the frequency of the maximum non-reciprocity (panel b) highlights the relevant temperature region in which the sample transition between the reciprocal and non-reciprocal states. Note, that the temperature was monitored at the mixing chamber plate to track the sample's electron and lattice temperature, both of which are thermalized to the mixing chamber plate temperature. \cite{Rosen2022} For reference, the time dependence of the 4-Kelvin stage temperature is also shown in panel c. No large temperature fluctuations which can result in gain fluctuation of the HEMT amplifier were observed. The temperature behaviour of the mixing chamber stage (MXC) is also shown. For the $R=100~\mu$m sample, the isolation sustains a sizable value up to 1 K at $f=0.6$ GHz. A close-up of the isolation is shown in Fig. S\ref{fig:R100}a. 

Figure S\ref{fig:R100}b shows the voltage pulses as a function of time detected for a short path (P1) and a long path (P2), respectively. As a result of the symmetric Carlin device structure, the ratio for P2/P1 is $2:1$. By subtracting P2 from P1, we reject the cross-talk between the ports and enhance the EMP signal (Fig. S\ref{fig:R100}c). From the Fourier transform of the curve, Figure S\ref{fig:R100}d, we clearly observe a peak around 0.6 GHz, in complete agreement with the resonant frequency observed in the microwave transmission measurement.

The $T$ dependent non-reciprocity decay rate in the $R=100 ~\mu$m sample also exhibits the variable range hoppping (VRH) behavior (Fig. S\ref{fig:R100_tau}). The fits to $\text{ln}(\beta T)$ vs. $T^{-1/2}$ yield the characteristic temperature $T_0 \sim $ $17$, $12$, and $10$ K above the threshold temperature $T_{th} \sim 0.4$ K for $P=-70$, $-45$, and $-20$ dBm, respectively.

\subsection{\label{sec:level2}Frequency and time domain response in the $R=25~\mu$m sample}

For the $R=25~\mu$m sample, we observed that the isolation behavior sustains up to $0.6$ K (Fig. S\ref{fig:R25}a). As shown in Figs. S\ref{fig:R25} b and c, we did not observe an obvious time delay between the two paths. The reason is that in this case the spatial width of the EMP wave packet is comparable with the perimeter of the sample, making delay between the two paths indistinguishable. Still, the Fast Fourier transform of the time-domain difference signal yields a frequency peak at 2 GHz that is comparable with the resonant frequency extracted from the frequency domain measurements. 

\section{\label{sec:level1}Power Dependent Microwave Transmission Measurements}

\begin{figure*}
\includegraphics[width=5.5in]{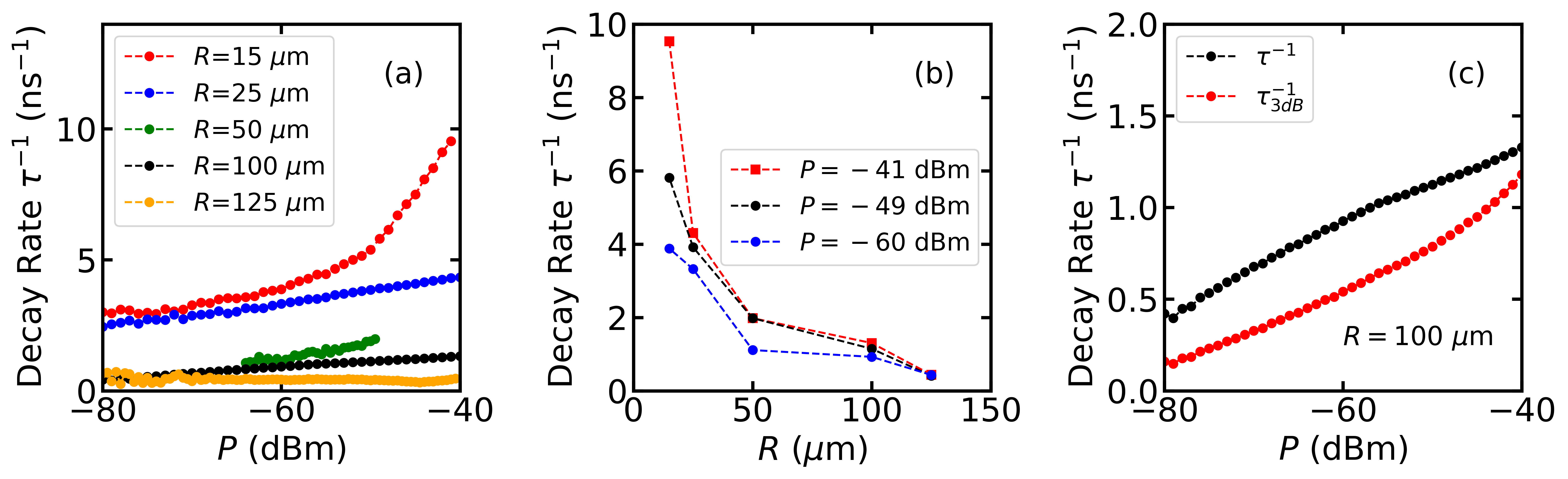}
\caption{\label{fig:Power_dependent_S}Microwave power and mesa size dependent decay rate. (a) The decay rate as a function of microwave power for five different samples. (b) The dependence of decay rate is plotted as a function of the mesa radius for the applied port power $P=-60$, $-49$, and $-41$ dBm, respectively. (c) Comparison of the non-reciprocity decay rate $\tau^{-1}$ and $\tau_{\text{3dB}}^{-1}$ derived from the FWHM of the Lorentzian fit to $S_{31}-S_{32}$ and the frequency width $\Delta f$ at $-3$dB of the $S_{31}-S_{32}$ peak, respectively. It has been demonstrated that the EMP decay rate can be estimated from the frequency domain data by measuring the quality factor.}
\end{figure*}

We obtained the non-reciprocity decay rate as a function of the excitation power and the device radius by fitting to the microwave transmission curves with a Lorentzian function. We find that the decay rate increases linearly as a function of the microwave power for $R = 25$, $50$, $100$, and $125$ $\mu$m devices (Fig. S\ref{fig:Power_dependent_S}a). For the smallest $R=15~\mu$m sample, the decay rate exponentially increases with $P$. By plotting the decay rate for different devices as a function of the input power (Fig. \ref{fig:Power_dependent_S}b), we observed a strong trend of increased decay rate as the device becomes smaller, implying that smaller devices normally lead to a faster plasmon decay with the increasing microwave power. We also confirm that the decay rate extracted from the FWHM of the Lorentzian fit to the non-reciprocity has the similar trend as the $-3$dB bandwidth fit does. 

\section{\label{sec:level1}Simulation with Circuit Model}
\begin{figure*}
\includegraphics[width=5.5in]{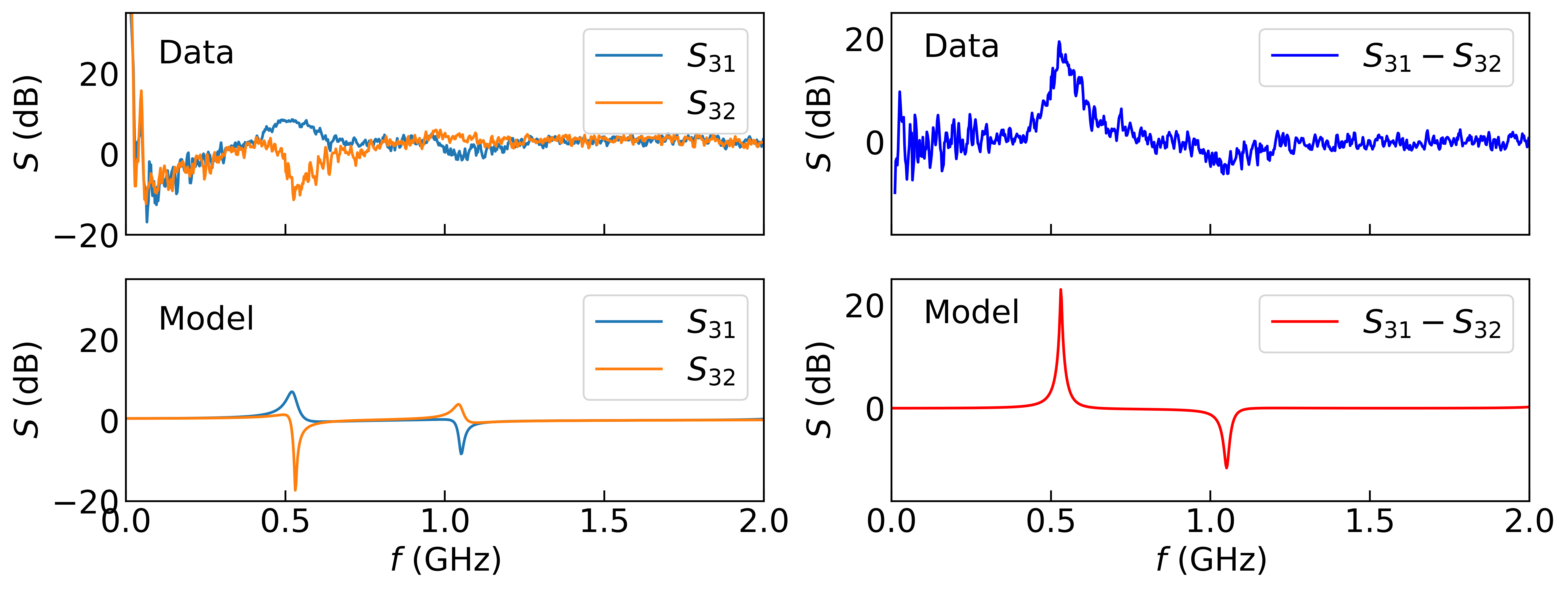}
\caption{\label{fig:Circuit_Model_S}The $S$-parameter characteristic is captured by the capacitively coupled circulator model. Experimental data (top panels) is compared with the simulation results (bottom panels) for the $R = 125~\mu$m sample. The model is simulated with the port-to-port parasitic capacitance $C_p=145$ fF, the edge channel capacitance $C_{edge}=24.7$ fF, the dissipation resistance $R=487~\Omega$, the Hall resistance $R_{xy}=25.8$ k$\Omega$, and the characteristic impedance $Z_0=50$ $\Omega$.}
\end{figure*}

In the Hall circulator model, the edge channel capacitance $C_{edge}$ consists of quantum capacitance, geometric capacitance, and the coupling capacitance between the 1D edge state and the charge puddles in the bulk. The admittance function of the EMP dynamics behaves like a transmission line with characteristic impedance $R_{xy}=1/\sigma_{xy}$ and channel capacitance $C_{edge}$. The dissipation resistance $R$ represents the resistive coupling between the edge channel and the surrounding circuit, primarily caused by the charge excitation in the bulk. The direct capacitive coupling between ports is accounted by the parasitic capacitance term $C_p$.\cite{Mahoney2017PRX} 

The circuit model qualitatively captures the non-reciprocity features in our sample. Figure S\ref{fig:Circuit_Model_S} shows the experimental and simulation results of signals for the $R=125~\mu$m sample. With the fixed parameters $R_{xy}$ and $Z_0$, $C_{edge}$ can be identified from the EMP resonance frequencies, while the dissipation resistance $R$ determines the quality factor of the resonance. The variation of $R$ obtained from the model is consistent with the scenario in which the dissipation increases as a result of the increasing microwave excitation. 

\FloatBarrier